\documentclass[review=false, screen, manuscript, nonacm, natbib=false]{acmart}
\usepackage{blindtext}
\usepackage{xpatch}
\makeatletter
\xpatchcmd{\ps@firstpagestyle}{Manuscript submitted to ACM}{}{\typeout{First patch succeeded}}{\typeout{first patch failed}}
\xpatchcmd{\ps@standardpagestyle}{Manuscript submitted to ACM}{}{\typeout{Second patch succeeded}}{\typeout{Second patch failed}}    \@ACM@manuscriptfalse
\makeatother

\renewcommand\footnotetextcopyrightpermission[1]{} 
\setcopyright{none}
\usepackage{xcolor}
\usepackage{xspace}

\usepackage[backend=biber, sorting=ynt, style=numeric, sortcites]{biblatex}
\newcommand{\prj}{\textsc{ConfD}\xspace}

\begin{document}

 \title{Analyzing Configuration Dependencies of  File Systems}

\author{
Tabassum Mahmud, Om Rameshwar Gatla, Duo Zhang, Carson Love, Ryan Bumann, Varun S Girimaji, Mai Zheng}

\affiliation{
\newline
    \institution{Department of Electrical and Computer Engineering, Iowa State University}
    \city{Ames}
    \state{IA}
    \country{USA}
}

\renewcommand{\shortauthors}{Mahmud et al.}

\begin{abstract}

File systems play an essential role in modern society for managing precious data. To meet diverse needs, they often support many configuration parameters.
Such flexibility comes at the price of additional complexity which can lead to subtle configuration-related issues.
To address this challenge, we study the
configuration-related issues 
of two major file systems (i.e., Ext4 and XFS) in depth, 
and  identify a prevalent pattern called multilevel configuration dependencies.
Based on the study, we build an extensible tool called \textsc{ConfD} to extract  the dependencies automatically, and create a set of plugins to address different configuration-related issues. 
Our experiments on Ext4, XFS and a modern copy-on-write file system (i.e., ZFS) show that
\prj was able to extract
 160 configuration dependencies for the file systems with a low false positive rate.
Moreover, the dependency-guided plugins can identify various configuration  issues  (e.g., mishandling of configurations,  regression test failures induced by valid configurations).
In addition, we also explore the applicability of \textsc{ConfD} on a popular storage engine (i.e., WiredTiger).
We hope that this comprehensive analysis of configuration dependencies of storage systems can shed light on addressing configuration-related challenges for the system community in general. 
\end{abstract}


\maketitle

\section{Introduction}
\label{sec:intro}

File systems (FS), such as Ext4 \cite{ext4} and XFS~\cite{XFS} on Linux-based  operating systems (OS) and NTFS~\cite{ntfs} on Windows OS, play an essential role in modern society.
They directly manage various files on desktops, laptops, and smartphones for numerous end users~\cite{Lu-FAST13-FS}. Moreover,
they often serve as the local storage backend for distributed storage systems (e.g., Lustre~\cite{lustre}, GFS~\cite{GoogleFS}, HopsFS~\cite{Hopfs}, MySQL NDB Cluster~\cite{mysqlndb}) to enable  storage management at scale.

To meet diverse needs, many file systems are designed with a wide range of configuration parameters controllable via  utilities~\cite{e2fsprogs,xfsprogs,btrfsbalance,fsckufs,zfsset,fsckminix,chkdsk,diskutil}, 
which enables   users to tune the systems with different tradeoffs.
For example,   Ext4 contains more than 85 configuration parameters which can be modified through a set of utilities called \texttt{e2fsprogs}~\cite{e2fsprogs}. The combination of the configuration parameters represents over $10^{37}$   configuration states~\cite{Carver}.

While configuration parameters have  improved the system flexibility, they introduce additional complexity for reliability. 
Subtle correctness issues often rely on specific parameters to trigger~\cite{confu, spex};  
consequently, they may elude  intensive testing  and affect  end users negatively.
For example, in December 2020, users of Windows OS observed that the checker utility of  NTFS (i.e., \texttt{ChkDsk}~\cite{chkdsk}) may  destroy NTFS on SSDs~\cite{hothardware20201220,borns20201218}. 
The incident turned out to be configuration-related: 
 two specific parameters must be satisfied to manifest the issue, including the `\texttt{/f}' parameter of \texttt{ChkDsk} and another (unnamed)  parameter in Windows OS~\cite{borns20201221}.

 \begin{figure}[tb]
	\centering
 \includegraphics[width=2.5in]{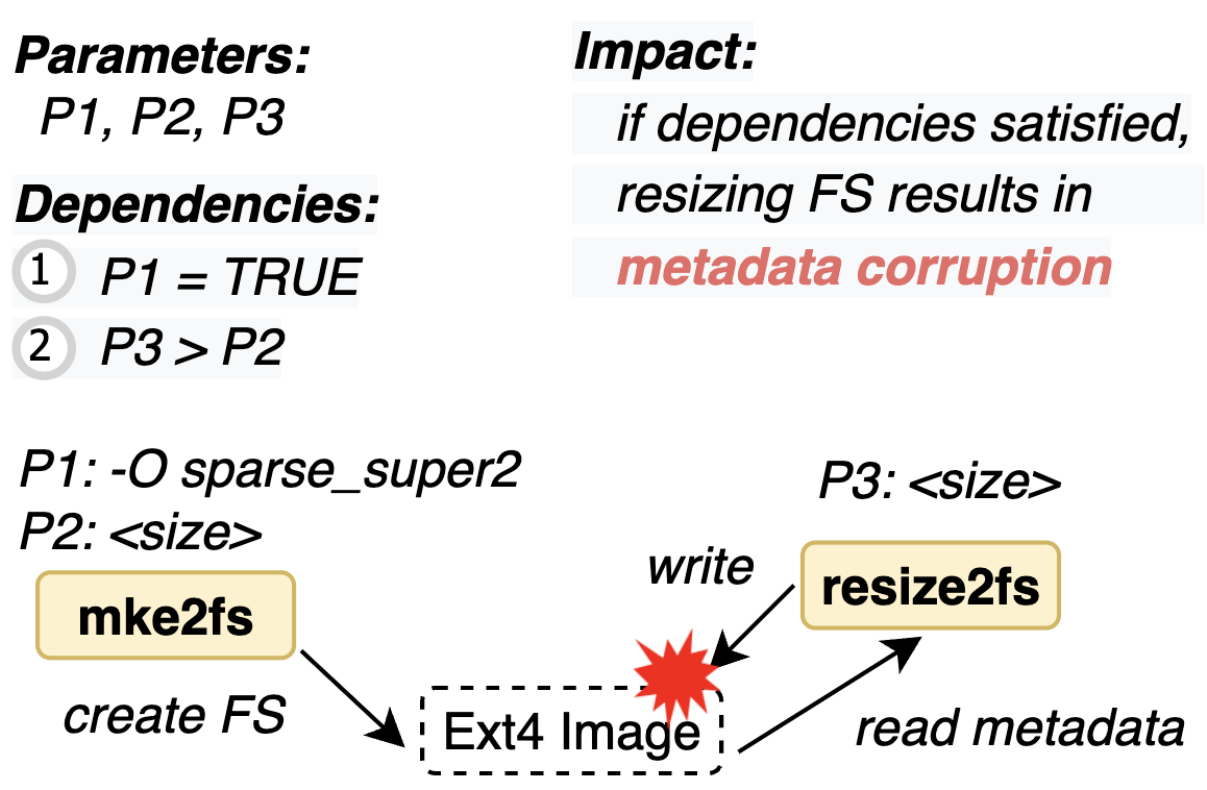}
	\caption{  {\bf A Configuration-Related Issue of Ext4}.  { When \texttt{sparse\_super2} feature is enabled and the \texttt{size} parameter of \texttt{resize2fs} is larger than the Ext4 size, expanding the file system results in metadata corruption.  
	}}
	\label{fig:resizebug}
\end{figure}

Similarly, Figure~\ref{fig:resizebug} shows another configuration-related issue involving Ext4 and its \texttt{mke2fs} and \texttt{resize2fs} utilities~\cite{e2fsprogs}.
Two conditions must hold to trigger the bug: 
(1) the \texttt{sparse\_super2} feature is enabled in Ext4 (via \texttt{mke2fs});
(2) the value of the \texttt{size} parameter of ~\texttt{resize2fs} must be larger than the size of Ext4 (i.e., expanding the file system). Once triggered, the bug will corrupt the Ext4 metadata with incorrect free blocks. {The root cause behind the issue was logical: with the specific configuration, the free block count of the last block group of Ext4 was calculated before adding new blocks for expansion.}

Due to the combinatorial explosion of configuration states and the substantial time needed to scrutinize a file system under each 
configuration state,  
it is practically impossible to exhaust all
states 
for thorough testing  today~\cite{
EMClarke-ModelCheckingAndTheStateExplosionProblem}.
Moreover, with more and more heterogeneous devices and advanced features 
 being introduced~\cite{Ext4_DAX,samsungsmartssd,trimsupport}, the  configuration states 
 are expected to grow. 
 Therefore,  effective methods to help improve configuration-related testing and identify critical configuration issues efficiently are much needed.

\subsection{Limitations of the State of the Art}
There are practical test suites  to ensure the correctness of file systems under different configurations (e.g., \texttt{xfstests}~\cite{xfstest}). 
Unfortunately, their coverage in terms of configuration is limited:
fewer than half of configuration parameters are used based on our study, which reflects the need for better tool support.
Also, configuration-related issues have   emerged in other software systems and have received much attention~\cite{cdep,ctest-OSDI20,Conferr-DSN08,spex,NavigatingMazeBioinfo-MikaelaCashman-ASE18}. But unfortunately, existing efforts mainly focus on relatively simple configuration issues  (e.g., typos~\cite{Conferr-DSN08}) within  one single application, which is  limited for addressing the file system configuration challenge involving multiple programs.

Please refer to \S\ref{sec:background} for more details.

\subsection{Our Efforts \& Contributions}
This paper presents one of the first steps to address the increasing configuration challenge of file systems. Inspired by a recent study~\cite{cdep} on configuration issues in Hadoop~\cite{Hadoop} and OpenStack~\cite{OpenStack}, we focus on \textit{configuration dependency}, which describes the dependent relations among configuration parameters~\cite{cdep}. Such dependency has been identified as a key source of  complexity caused problems, and capturing the dependency is essential for improving  configuration design and tooling~\cite{cdep,TianyinXu-AnHCIViewofConfigurationProblemsArXiv16,spex}.

While the basic concept of configuration dependency  
has been proposed in the literature (see \S\ref{sec:background}),
the understanding of specific dependency patterns and  implications  in the context of  file  systems is still limited. 
Therefore, we first conducted an empirical study on 78 configuration-related issues in two major file systems (i.e., Ext4 and XFS).
By scrutinizing   real-world   bugs and the relevant source code, 
we answer one important question: What critical configuration dependencies exist in file systems? 

Our study reveals a prevalent pattern called \textit{multilevel configuration dependencies}.
Besides the relatively simple configuration constraints (e.g., value range~\cite{spex}),
there are implicit dependencies among parameters from different utilities of a file system.
The majority  (96.2\%) of  issues in our dataset requires meeting 
such deep configuration dependencies to manifest.
Interestingly, the  workloads applied to the file system do not have to be configuration-specific: 
71.8\% issues only involve generic file system operations. 

Based on the study, we built an extensible framework called \textsc{ConfD} to extract the multilevel configuration dependencies automatically and leverage dependency-guided   configuration states for further analysis. 
One key challenge is how to establish the correlation between parameters specified through different utilities which have different ways of  configuration handling.
We address the challenge by metadata-assisted taint analysis, which leverages the fact that all utilities of a given file system share the same  metadata structures.  
Moreover, based on the   dependencies extracted,  we created a set of plugins to help address  configuration-related issues in file systems from different angles.

 Our experiments show that
\textsc{ConfD}  can extract 154 different configuration dependencies 
with a low false positive rate (8.4\%) for  Ext4 and XFS. 
Moreover, with the dependency guidance, 
the \textsc{ConfD} plugins can  
identify various configuration-related issues, including inaccurate documentations, configuration handling issues, and regression test failures induced by valid configurations.

{This article is an extension of our previous work \cite{mahmud2023confd}. In this version, we explored further the applicability of \textsc{ConfD} on two new storage systems including a modern copy-on-write file system (i.e., ZFS~\cite{zfs}), and a popular NoSQL storage engine (i.e.,  WiredTiger~\cite{wt}) and extracted the configuration dependencies in the new context. Moreover, based on the dependencies extracted from ZFS, we introduced a new plugin to help enhance the regression test suite of ZFS.
In addition, we  optimized the plugin for regression testing of Ext4 and XFS  which reduces the false positive rate  by 68\% on average. 
Through this comprehensive analysis of configuration dependencies of storage systems, 
we hope that our effort can shed light on addressing configuration-related challenges for practical systems in general. 

In summary, this paper makes the following contributions:
\begin{itemize}
    \item Deriving a taxonomy of critical  configuration dependencies of file systems based on  real-world  issues from two major Linux file systems (i.e., Ext4, XFS).
    \item Building the \textsc{ConfD} prototype \footnote{\textsc{ConfD} is on \url{https://github.com/data-storage-lab/ConfD}}
    to  extract configuration dependencies from three file systems (i.e., Ext4, XFS, ZFS) and one database engine (i.e., WiredTiger), and expose relevant issues in storage systems. 
    \item Integrating  with multiple practical tools (e.g.,fault injector~\cite{OmFAST18}, fuzzer~\cite{hydra-2019},  regression test suites~\cite{xfstest,e2fsprogs-test}) to improve their  configuration coverage and effectiveness.
\end{itemize}
}

The rest of the paper is organized as follows: \S\ref{sec:background} introduces the background and related work; \S\ref{sec:bugstudy} presents the empirical study and findings; \S\ref{sec:design-extract} describes the  \textsc{ConfD} framework;
\S\ref{sec:results} shows experimental results; \S\ref{sec:discussion} discusses our observations, limitations and potential extensions;   \S\ref{sec:conclusion} concludes the paper.
\section{Background \& Related Work}
\label{sec:background}

\subsection{Background}

\begin{figure}[tb]
	\centering
    \includegraphics[width=3.3in]{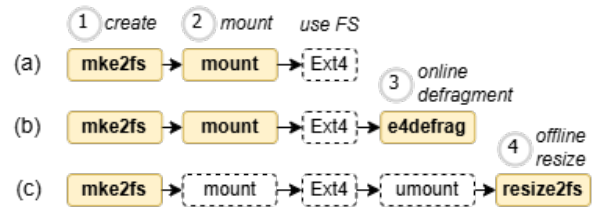}
   \vspace{-0.1in}
	\caption{  {\bf Methods of Configuring File Systems}. {This figure shows 
	four typical stages to configure a file system: (a) at creation (e.g., \texttt{mke2fs}) or mount time (\texttt{mount}) before usage; (b) via  online utilities (e.g., \texttt{e4defrag}); (c) via offline utilities. }} 
	\label{fig:usage}
\end{figure}

\noindent
\textbf{File System Configurations.}   
The configuration methods of file systems are different from that of many applications, {which makes the problem arguably more challenging}. As shown in Figure~\ref{fig:usage}, a typical file system may be configured through a set of utilities at four different stages:
    \vspace{-0.05in}
\begin{itemize}
    \item \textbf{Create}. When creating file systems,  the \texttt{mkfs} utility (e.g., \texttt{mke2fs} for Ext4) generates the initial configurations.
    \item \textbf{Mount}. When mounting file systems, certain configurations can be specified via \texttt{mount} (e.g., `\texttt{-o dax}' to enable the Direct Access or DAX feature~\cite{Ext4_DAX}). 
     \item \textbf{Online}. Many  utilities can change the  configurations of a mounted file system directly by modifying the metadata online (e.g., Ext4 defragmenter \texttt{e4defrag}~\cite{e4defrag},  Windows NTFS checker   \texttt{ChkDsk}~\cite{chkdsk}).       
     \item \textbf{Offline}. Offline utilities can also modify file system images and change  the configurations   (e.g.,   \texttt{resize2fs}~\cite{resize2fs},   \texttt{e2fsck}~\cite{e2fsck})
     
\end{itemize}

 Note that all the  utilities  have different configuration parameters to control their own behaviors, which will eventually affect the file system state. 
{Moreover, the configuration parameters may affect the behavior of the file system long after the FS image is created, and some configurations cannot be changed later. Also,}
  the validation of parameters may occur at both user level and kernel level.
 For example, the `\texttt{-O inline\_data}' parameter of \texttt{mke2fs} and the `\texttt{-o dax}' of \texttt{mount} are further validated in the \texttt{ext4\_fill\_super}  function of Ext4.
Therefore, we believe it is necessary to consider the file system itself as well as all the associated utilities as an \textit{FS ecosystem} to address the configuration challenge.
For simplicity, we call the  file system  and utilities as \textit{components} within the FS ecosystem.

The multi-stage configuration method  is common among   file systems. As listed in Table~\ref{tab:manyFSandUtilities}, many popular file systems follow  similar modular designs and can be configured via different utilities at different stages.
Therefore, we believe that the multi-component configuration challenge is general.

\begin{table}[t]
	\small
	\begin{center}
		\begin{tabular}{ c | c | c | c | c }
	\textbf{FS (OS)}  & \multicolumn{4}{c}{\textbf{Four  Stages of Configuration}}  \\
			\cline{2-5}
			\textbf{ } & \textbf{Create} & \textbf{Mount} &  \textbf{Online} & \textbf{Offline} \\ 
			\hline
			{Ext4 (Linux)} & \cite{mke2fs} & \cite{mount} & \cite{e4defrag}, \cite{resize2fs} & \cite{e2fsck}, \cite{resize2fs}  \\
			\hline
			{XFS (Linux)} & \cite{mkfsxfs} & \cite{mount} & \cite{xfsfsr}, \cite{xfsgrowfs} & \cite{xfsadmin}, \cite{xfsrepair}  \\
			\hline
			{BtrFS (Linux)} & \cite{mkfsbtrfs} & \cite{mount} & \cite{btrfsbalance}, \cite{btrfsscrub}  & \cite{btrfscheck} \\
			\hline
			{UFS (FreeBSD)} & \cite{newfs} & \cite{mountfreebsd} & \cite{growfs}, \cite{restore} & \cite{dump}, \cite{fsckufs}  \\
			\hline
			{ZFS (FreeBSD)} & \cite{zfs-create} & \cite{zfs-mount} & \cite{zfs-rollback}, \cite{zfsset} & \cite{zfs-destroy} \\
			\hline
			{NTFS (Windows)} & \cite{format} & \cite{mountvol} & \cite{chkdsk}, \cite{defrag}  & \cite{chkdsk}, \cite{shrink} \\
			\hline
			{APFS (MacOS)} & \cite{diskutil} & 
			\cite{mountapfs} &
			\cite{diskutil}  & \cite{diskutil}, \cite{fsckapfs} \\
			\hline
		\end{tabular}
	\end{center}
	\caption{ {\bf Examples of configuration methods for different file systems}. The last four columns  list example  utilities that can affect the file system configuration states.
	}
	\label{tab:manyFSandUtilities}
\end{table}

\smallskip
\noindent
\textbf{FS Test Suites.}  
Practical test suites have been created to ensure the correctness of file systems under various configurations. Unfortunately, 
due to the complexity  of configurations, their coverage in terms of configuration is limited. As shown in Table~\ref{tab:testsuites},
fewer than half of configuration parameters are used in  the standard test suites of Linux file systems (i.e., \texttt{xfstests}~\cite{xfstest}, \texttt{e2fsprogs/tests}~\cite{e2fsprogs-test}) based on our study.
Since each parameter may have a wide range of values representing different states, the total number of 
missed configuration states is much more than the number of unused  parameters, which implies the need for better tool support.

\begin{table}[t]
	\small
	\begin{center}
		\begin{tabular}{ c | c | c | c  }
			\textbf{Test} & \textbf{Target} &  \multicolumn{2}{c}{\textbf{\# of Conf. Param.}}  \\
			\cline{3-4}
			\textbf{Suite} & \textbf{Software} & \textbf{Total} & \textbf{Used}  \\
			\hline
			\texttt{xfstests/ext4} & 	{Ext4} & $>$85 & 29 ($<$ 34.1\%)   \\
			\hline
			\texttt{e2fsprogs} & \texttt{e2fsck} & $>$35 & 6 ($<$ 17.1\%)  \\
		     	\texttt{/tests} & \texttt{resize2fs}  & $>$15 & 7 ($<$ 46.7\%)  \\
			\hline
		\end{tabular}
	\end{center}
 
	\caption{ {\bf Configuration Coverage of Test Suites.} 
	}
 	 \vspace{-0.1in}
	\label{tab:testsuites}
\end{table}

\smallskip
\noindent
\textbf{Configuration Constraints \& Dependencies.}
 Configuration \emph{constraints}  specify the configuration requirements (e.g., data type,  value range) of software~\cite{spex}. Intuitively, such information can help identify important configuration states, and it has proved to be effective for addressing  configuration-related  issues in a wide range of applications~\cite{cdep,Conferr-DSN08,spex,AnEmpiricalStudy, ConfigEverywhere-DongpuJin-ICSE14}. 
Configuration \emph{dependency} is one special type of constraint describing the dependent correlation among parameters~\cite{cdep,spex}, which has shown recently to be critical for addressing complex configuration issues in cloud systems~\cite{cdep}.
For simplicity, we use constraints and dependencies interchangeably in the rest of the paper.
Note that although the basic concepts have been proposed, 
there is limited understanding of them in the context of file systems. 
This paper attempts to fill the gap.

\vspace{-0.05in}
\subsection{Related Work}
\label{sec:related}
\vspace{-0.05in}

\smallskip
\noindent
{\bf Analysis of Software Configurations.} 
Configuration issues have been  studied in many software applications~\cite{cdep,ctest-OSDI20,confu,Conferr-DSN08,spex,ConfigEverywhere-DongpuJin-ICSE14,NavigatingMazeBioinfo-MikaelaCashman-ASE18,ArielRabkin-ICSE11}. 
For example, ConfErr~\cite{Conferr-DSN08} manipulates parameters to emulate human errors;  
Ctests~\cite{ctest-OSDI20} detects failure-inducing configuration changes.
In general, these works do not analyze deep   dependencies within the  software.
The closest work is {cDEP}~\cite{cdep},
which notably observes \textit{inter-component dependencies} in  Hadoop~\cite{Hadoop} and OpenStack~\cite{OpenStack}. Unfortunately, their solution is largely inapplicable for file systems. 
This is because their target components share   configuration specifications (e.g., XML)  and libraries~\cite{ApacheCommonsConfiguraitons2}, which makes them equivalent to one single  program in terms of configuration. 
In contrast, 
the configuration dependencies in file systems may cross different programs and the user-kernel boundary, which requires  non-trivial mechanisms to extract.
In addition, cDEP relies on a Java   framework~\cite{soot} 
which cannot handle C-based file systems. 

\smallskip
\noindent
{\bf Reliability of File Systems.} 
Great efforts have been made to improve the reliability of 
file systems~\cite{bornholt2016specifying,recon12,hydra-2019,Changwoo-SOSP15-CrosscheckingFS,iron05} 
and  their  utilities~\cite{OmFAST18,Om-TOS18,SQCK,spiffy,SWIFT}.
For example, 
Prabhakaran et al.~\cite{iron05} apply fault injection to analyze the failure policies of  file
systems and propose improved designs based on the IRON taxonomy; 
Xu et al.~\cite{xu2019fuzzing} and Kim et al.~\cite{hydra-2019} use fuzzing to detect file system bugs;  
SQCK ~\cite{SQCK} and RFSCK~\cite{OmFAST18} improve the checker utilities of file systems to avoid inaccurate fixes. 
 While effective for their original  goals, these works do not consider multi-component configuration issues. 
 On the other hand, the configuration dependencies from this work may be integrated with these existing efforts to improve their coverage (see \S\ref{sec:plugins}).  
 Therefore, we view them as complementary. 
 
\smallskip
\noindent
{\bf Configuration Management Tools.} Faced by the increasing  challenge, practitioners have created dedicated frameworks for configuration management 
~\cite{CFEngine,Yadan2019Hydra}. 
For example,  
Facebook HYDRA~\cite{Yadan2019Hydra} supports managing hierarchical configurations elegantly. 
While helpful for developing new applications, refactoring FS ecosystems to leverage such frameworks would require substantial efforts (if possible at all).
Notably, the framework supports running a program with different compositions of configurations automatically. Nevertheless,  since it does not understand configuration dependencies, it may generate many invalid configuration states (see \S\ref{sec:state-generation-vs-fb-hydra}). This work aims to address such limitations.

\section{Configuration Dependencies in File Systems}
\label{sec:bugstudy}
  \vspace{-0.05in}

In this section, we present a study on the Ext4 and XFS ecosystems to understand the potential patterns of configuration issues and guide the design of solutions. 
We discuss the methodology and  findings in \S\ref{sec:methodology}  and \S\ref{sec:findings}, respectively. 

\begin{table*}[t]
	\small
	\begin{center}
		\begin{tabular}{ l | l | c | c | c | c | c }
	\multicolumn{2}{l|}{\textbf{FS Usage Scenarios}} & \textbf{Description} & \textbf{ \# of}  & \multicolumn{3}{c}{\textbf{Multilevel Config. Dependencies}}\\
         \cline{5-7}
		
	\multicolumn{2}{l|}{	 {(key configuration utilities are in bold)}} &   &  \textbf{ Bug} &  \textbf{SD}  &   \textbf{CPD}  &  \textbf{CCD}    \\
					\hline
1 & \textbf{mke2fs} - \textbf{mount} - Ext4 & create \&	mount an Ext4 to use & 13  & 13 (100\%) & 1 (7.7\%) & 13 (100\%) \\
			\hline
2 & \textbf{mke2fs} - \textbf{mount} - Ext4 - \textbf{e4defrag} & online defragmentation & 1 &  1 (100\%)  & -- & -- \\
			\hline
3 & 	\textbf{mke2fs} - mount - Ext4 - umount - \textbf{{resize2fs}}  &  resize an umounted Ext4  & 17 &  17 (100\%)  & -- & 17 (100\%)\\
			\hline
4 &  \textbf{mke2fs} - mount - Ext4 - umount - \textbf{{e2fsck}}  &  check  Ext4 \& fix inconsistencies   & 36 &  36 (100\%) & 4 (11.1\%) & 34 (94.4\%)\\
\hline

5 &  \textbf{mkfs.xfs} - \textbf{mount} - XFS & create \&	mount an XFS to use & 5&5 (100\%)& 2 (40\%)& 5 (100\%)\\
 \hline
6 &  \textbf{mkfs.xfs} - mount - XFS - umount - \textbf{xfs\_repair} & check XFS   \& fix  inconsistencies &6 &6 (100\%) &1 (16.7\%) &6 (100\%)\\
 \hline

\multicolumn{3}{r|}{\textbf{Total}}  & 78 &  78  (100\%) & 8 (10.3\%) & 75 (96.2\%)  \\
\hline
		\end{tabular}
	\end{center}
	\caption{ {\bf Distribution of Configuration Bugs in Six Scenarios.} This table shows  the distribution of 78 configuration bugs in six typical usage  scenarios of file system. The last three columns shows the percentages of bug cases that involve Self-Dependency (SD), Cross-Parameter Dependency (CPD), and Cross-Component Dependency (CCD), respectively. 
}
	\label{tab:bugstudy}
\end{table*}

\begin{table*}[t]
	\small
	\begin{center}
		\begin{tabular}{ c | c | c | c | c  }
		\multicolumn{2}{c|}{	\textbf{Multilevel Config. Dependencies}} & \textbf{Description} & \textbf{Observed?}  &  \textbf{Count}  \\
          \hline
Self Dependency	  &  Data Type  & parameter $P$ must be of a specific data type (e.g., integer)   &   Y  &  44\\
					\cline{2-5}
   (SD)        & Value Range & $P$  must be within a specific  {value range} (e.g., $P$  $<$ 4096)  & Y & 41\\
			\hline
Cross-Parameter      & Control  &  $P1$  of $C1$ can   be enabled iff   $P2$  of $C1$  is enabled/disabled & Y & 5 \\
					\cline{2-5}
  Dependency      &  Value    &  $P1$'s  value depends on   $P2$ 's value (e.g., $P1 < P2$)   &  N & -- \\
             \cline{2-5}
     (CPD) &    Behavioral &     component $C1$'s behavior depends on $P1$ and $P2$ of component $C1$ & Y & 1 \\ 

			\hline
Cross-Component      & Control    & $P1$  of $C1$ can   be enabled iff   $P2$  of $C2$  is enabled/disabled &  Y&  1 \\
					\cline{2-5}
 Dependency    & Value   &$P1$'s  value depends on  $P2$ from another component   &  Y & 1\\
 \cline{2-5}
 (CCD)  & Behavioral   &  component $C1$'s behavior  depends on  $P2$ of  $C2$ &  Y&  75\\
 \hline
 \multicolumn{3}{r|}{\textbf{Total}} & 7/8  & 168  \\

\hline
		\end{tabular}
	\end{center}
	\caption{ {\bf Multilevel Configuration Dependencies.} This table describes the multilevel configuration dependencies observed. $Pn$ means parameter, $Cn$ means component. The last column shows the count of each sub-category of dependency observed. 
}
	\label{tab:dependencies}
\end{table*}

\subsection{Methodology}
\label{sec:methodology}

Our dateset includes two parts: (1) the source code of Ext4 and XFS and
seven important utilities including \texttt{mke2fs}, \texttt{mount}, \texttt{e4defrag}, \texttt{resize2fs}, \texttt{e2fsck}, \texttt{mkfs.xfs}, and \texttt{xfs\_repair}, which are described in Table~\ref{tab:bugstudy};
(2) a set of 78 configuration-related bug patches for the two FS ecosystems, 
which are collected from the commit histories of their source code repositories via a combination of keyword search (e.g., ‘configuration’, ‘parameter’, ‘option’), random sampling, and manual validation. 
Note that the patch collection method is inspired by previous studies of real-world bugs~\cite{Lu-FAST13-FS,lu2008learning,Duo-SYSTOR21}. While time-consuming, it has proved to be valuable for driving system improvements~\cite{Lu-FAST13-FS,lu2008learning}.
On the other hand, similar to previous studies, the findings of our study should be interpreted with the method in
mind. For example, the 78 patches only represent a subset of issues that have been triggered and fixed; there are likely  other configuration-related issues  not yet discovered (see \S\ref{sec:discussion} for further discussion). 

\vspace{-0.05in}
\subsection{Findings}
\label{sec:findings}
Based on the   dataset,  we analyzed  each patch 
and the relevant source code in depth to understand the logic,
which enables us to identify the configuration usage scenarios  as well as  configuration constraints that are critical. We summarize our findings in Table~\ref{tab:bugstudy} and Table~\ref{tab:dependencies} and discuss them below. 

\vspace{0.1in}
\noindent
\textbf{Finding \#1:} \textit{The majority of cases (96.2\%) involve critical  parameters from more than one component}.  
The first column of Table~\ref{tab:bugstudy} shows six typical usage scenarios of file systems which cover all bug cases in our dataset (78 in total). 96.2\% of the bug cases require specific parameters from at least two key utilities (i.e., the utilities in bold in each usage scenario) to manifest.
 
This reflects the complexity of the  issues  and suggests that we cannot only consider one single component.

\noindent
\textbf{Finding \#2:} \textit{There is a hierarchy of configuration dependencies}.
We classify the configuration constraints derived from our dataset into three major categories as follows:

\begin{itemize}

  \vspace{-0.05in}
    \item \textbf{Self Dependency (SD)} means individual parameters  must satisfy their own  constraints (e.g., data type or value range). For example,  the \texttt{blocksize} parameter of \texttt{mke2fs} has a value range of 1024 - 65536 and must be a power of 2. 
    
    \item \textbf{Cross-Parameter Dependency (CPD)} means multiple parameters of the same component must satisfy relative relation constraints (e.g., two \texttt{mke2fs} parameters \texttt{meta\_bg} and \texttt{resize\_inode} cannot be used together). 
       
    \item \textbf{Cross-Component Dependency (CCD)} means the parameters or behaviors of  one component  depend on the  parameters of another component. 
    Both dependencies   in Figure~\ref{fig:resizebug} belong to this category becasue they involve parameters of \texttt{mke2fs} and the (buggy) behavior of \texttt{resize2fs} depend on them.

\end{itemize}

As summarized in Table~\ref{tab:dependencies}, 
each major category may contain a couple of sub-categories which describe more specific constraints. 
 Together, these constraints form a hierarchy which we call \textit{multilevel configuration dependencies}.
 Note that we only observed 7 out of 8 sub-categories  in the dataset. We include the  unseen ``Value'' sub-category in CPD based on the literature~\cite{spex} for completeness.

Moreover, among all the dependencies, there is a subset which directly contribute to the manifestation of the bugs in our dataset: the relevant parameters are explicitly mentioned in the bug patches, and modifications to the corresponding functionalities are needed to fix the bugs (i.e., they are related to the root causes). We call this subset of dependencies as \textit{critical dependencies}.
The count of the critical dependencies for each sub-category is shown in the last column of  Table~\ref{tab:dependencies}. 
We are able to derive 168 critical dependencies  manually  in total, which is larger than the number of bug cases. This is because multiple critical dependencies may be needed to trigger a bug. For example, both dependencies in Figure~\ref{fig:resizebug} are critical dependencies for this bug case.

As shown in the last three columns of Table~\ref{tab:bugstudy}, SD  and CCD  are almost always involved in all scenarios (100\% and 96.2\% respectively), while CPD is non-negligible (10.3\%).
This is because SD represents relatively simple constraints which always need to be satisfied first  {to make the target component work} (e.g.,  correct spelling). 
SD is relatively easy to check and has been the focus of previous work~\cite{Conferr-DSN08}.
However, this does not mean that 100\% of the bugs could be avoided if SD is checked or satisfied. 
For example, a bug related to both the \texttt{bigalloc} and \texttt{extent} parameters (i.e., there is a CPD involved) may still occur even if the two parameters are spelled correctly.
In other words, only considering  simple constraints of individual parameters is not enough. 

Interestingly, we observed both CPD and CCD between the DAX feature and other seemingly irrelevant configurations. 
 In one case, a corruption was triggered when
`\texttt{-O inline\_data}' was used in \texttt{mke2fs} and the image was mounted with `\texttt{-o dax}' subsequently. 
In another case, the DAX  feature conflicted with the  `\texttt{has\_journal}' configuration, which may lead to corruptions when changing the journaling mode online.   Such unexpected dependencies implies the complexity of adding the DAX support to the Linux kernel.

\smallskip
\noindent
{\textbf{Finding \#3:} \textit{Configuration parameters are handled in heterogeneous ways in an FS ecosystem}.} 
We identified four major sources of heterogeneity in  FS configurations. First, different parameters may be mapped to different types of variables in the code. For example,   the parameters of Ext4 may be stored  in (at least) four different ways including (i) a local variable, (ii) a global variable, (iii) a bit in a  bitmap   accessed via bit operations, and (iv) directly in the superblock.    
Second, within the superblock, parameters may be kept either in one single field (e.g., \texttt{ s\_log\_block\_size}) or as one member of a compound field. Third, parameters can be   loaded from the superblock either directly or through library calls. Lastly, different components may use different functions for handling configurations  (e.g., \texttt{resize2fs}  uses the ``main'' function, while \texttt{mke2fs} invokes a special function called ``PRS'').
Such heterogeneity makes previous solutions mostly inapplicable.

\smallskip
\noindent
 \textbf{Finding \#4:} \textit{The majority of cases (71.8\%)  do not require configuration-specific workloads to manifest}. 
Interestingly, despite the complexity, many bugs can be triggered without applying configuration-specific workloads. This  suggests that we may re-use existing efforts on stressing file systems~\cite{e2fsprogs-test,xfstest} to analyze configuration-related issues effectively.

\section{Extracting \& Using Multilevel Configuration Dependencies}
\label{sec:design-extract}

\begin{figure}[tb]
	\centering
    \includegraphics[width=3.3in]{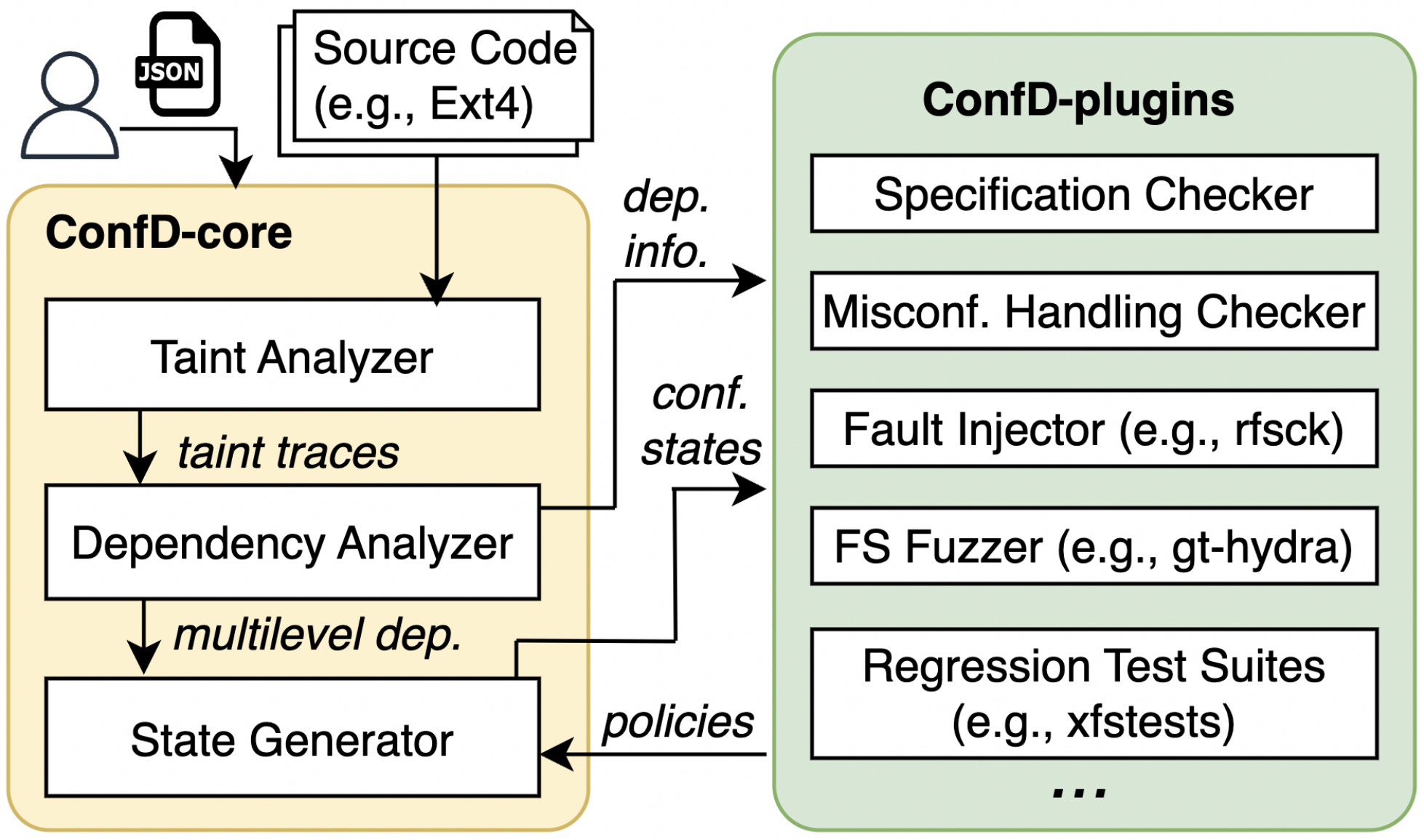}
    \vspace{-0.1in}
	\caption{  {\bf Overview of \textsc{ConfD}}. {There are two parts: (1) ConfD-core (yellow) for extracting  configuration dependencies and generating critical states; (2) ConfD-plugins (green) for detecting various configuration-related issues.} 
	}
	\label{fig:overview}
\end{figure}

 Based on the study, we built an extensible framework called \textsc{ConfD} to leverage the  dependency information to address configuration-related issues. 
As shown in Figure~\ref{fig:overview},   \textsc{ConfD}   consists of two main  parts:  (1) \textit{ConfD-core} (yellow box) for extracting multilevel  configuration dependencies and generating critical configuration states, which further contains three sub-modules (i.e., \textit{Taint Analyzer}, \textit{Dependency Analyzer}, and \textit{State Generator});
(2) \textit{ConfD-plugins} (green box) for detecting various  configuration-related issues based on the generated configuration states. We elaborate on the two parts in subsections \S\ref{sec:extraction} and  \S\ref{sec:leverage} respectively. Moreover, we further discuss optimizations for reducing false positive rates of plugins (\S\ref{sec:reducefalsepositives}) and  extensions for databases (\S\ref{sec:DBextension})  in the rest of the section.
 
 \subsection{Extracting Configuration Dependencies for File Systems}
 \label{sec:extraction}

\subsubsection{Metadata-assisted Taint Analysis}
\label{sec:derive}

As the first step, the \textit{Taint Analyzer} of  \textsc{ConfD} performs metadata-assisted taint analysis and generates taint traces to capture the propagation flow of configuration parameters in the target FS ecosystem.

 It takes the source code of the target system as input, 
and uses the LLVM compiler infrastructure~\cite{LLVM} 
to generate intermediate representation (IR) of the source code.
It then tracks the propagation of each configuration parameter  along the data-flow paths in IR based on the classic taint analysis algorithm~\cite{DrChecker}. 
We maintain  a set to keep the initial configuration variables and any variables  derived from the initial configuration variables while traversing the IR. When a new variable is added to the set, we add the corresponding IR instruction to the taint trace. 
We maintain a mapping between each configuration parameter and  the variables derived from it to enable tracking if a variable may be derived from multiple parameters, which is essential for establishing the correlation across parameters.
Our taint analysis is context-sensitive and can handle both intra-procedural and inter-procedural analysis. Context-sensitivity is important  for inter-procedural analysis because one function can be called from different contexts, which is also crucial for deriving accurate dependency across different taint traces (\S\ref{sec:dependencyAnalysis}).

One unique challenge we encounter is how to establish the mapping between  parameters of different components of the FS ecosystem.
As mentioned in \S\ref{sec:findings},
the components in the FS ecosystem tend to load configurations in different ways and process equivalent FS information using different variables or functions. We address this challenge based on one key observation: all  components need to access the same FS metadata structures. We can leverage  shared metadata structures   to connect relevant  parameters of different components. 

More specifically, the parameter values relevant to the FS configuration are (eventually) stored in the superblock structure of the file system. 
For example, the parameter \texttt{-I inode-size} from \texttt{mke2fs} is stored as the {27th} member of   the superblock (\texttt{s\_inode\_size}). When another component (e.g., \texttt{e2fsck}) loads the  \texttt{s\_inode\_size}   from the superblock to access it, it is essentially dependent on the \texttt{-I inode-size} parameter of  \texttt{mke2fs}.
We map the \texttt{mke2fs} parameter values to  relevant superblock fields by tracking where the parameter value is being written in the superblock. Similarly, the accesses to the superblock in other components are also tracked. Based on the mapping to the same superblock fields (e.g., \texttt{s\_inode\_size}), we can establish the connection between  taint traces from different components. 

{Note that since \textsc{ConfD} implements the taint analysis at the LLVM IR level, any file system that can be compiled to LLVM IR may benefit from it for configuration dependency analysis. The current prototype uses the Clang frontend of LLVM which supports C/C++/Objective-C languages~\cite{LLVM}.}

\subsubsection{Multilevel Dependency Analysis}
\label{sec:dependencyAnalysis}
Given the taint trace of every configuration parameter, the \textit{Dependency Analyzer} further analyzes the potential correlations between   parameters based on the multilevel dependencies  derived from our study  (\S\ref{sec:bugstudy}). 

Specifically,  
the self-dependency (SD) for each parameter is derived from their individual taint traces based on the data type and value range of the  variables. We also examine the  error statement immediately following a range check based on the observation that an error statement may indicate an invalid range. 
For CPD and CCD, we  compare taint traces of multiple parameters.
If there are common lines   (which are context-sensitive), we consider them to be dependent. 

Moreover, 
after getting the dependent parameters, we also leverage the subsequent error statements to further analyze the specific types of dependency (e.g.,  should be enabled or disabled together).  For example, the two parameters  \texttt{resize\_inode} and \texttt{meta\_bg} from \texttt{mke2fs}  cannot be enabled together, so there must be a common error statement immediately following the condition check shared by the two taint traces.

All of the extracted dependencies are stored in the JSON format~\cite{JSON} to describe both the parameters and the corresponding dependent relations concisely.

\subsubsection{Dependency-guided State Generation}
\label{sec:stategenerator}

With the dependency information, the \textit{State Generator} generates concrete configuration states for further analysis.  Instead of randomly generating combinations of configurations which may easily lead to useless states (\S\ref{sec:state-generation-vs-fb-hydra}), 
 it leverages the 
extracted multilevel  dependencies to generate states selectively.
 
Specifically, the \textit{State Generator}  uses a tree structure to maintain different configuration states. The root of the tree represents a default configuration state, and each child node on the tree represents a configuration state with exactly one modification made from its parent.
The module operates similar to a Depth First Search (DFS) on a tree, except it leverages the dependency information to guide which children nodes are worth pursuing. For example, given the cross-parameter dependency (CPD) between the \texttt{bigalloc} and \texttt{blocksize} parameters of \texttt{mke2fs}, if the current node modifies  \texttt{bigalloc}, then the child node to consider will be a state with a modification to \texttt{blocksize}.

Moreover, the module has a number of options that allow for tuning based on  needs. The first option is `depth',  which dictates how deep the DFS is allowed to go.  A larger value results in a greater number of  states being generated. The default `depth' is 3 which worked well in our experiments. 
Another option is the `policy' under which the State Generator operates. There are two basic policies as follows:

\smallskip
\noindent
   \textbf{Following Dependency}. Under this policy, we always honor the extracted multilevel dependencies when creating a configuration state. For example, \texttt{sparse\_super} should always be enabled if \texttt{resize\_inode} is enabled for~\texttt{mke2fs} according to the multilevel dependency, so the module may generate a   state with both parameters (i.e., `\texttt{mk2fs -O resize\_inode,sparse\_super}').
 
Essentially, this policy only generates  \textit{valid} configuration states involving critical  parameters for the target FS ecosystem, which is  the basic requirement for running many FS applications or tools properly.
Note that this policy is consistent with recent work on testing configuration changes which shows that \textit{valid} configuration changes may induce production failures~\cite{ctest-OSDI20}.

\smallskip
\noindent
   \textbf{Violating Dependency}. Under this policy, we intentionally violate   the multilevel dependencies when creating a configuration state. For example,
   the \texttt{resize\_inode}  and  \texttt{sparse\_super} parameters of \texttt{mke2fs} have  a cross-parameter dependency (CPD):  \texttt{sparse\_super} must be enabled if we want to enable \texttt{resize\_inode}. To violate the CPD, the module may intentionally generate a state which disables the \texttt{sparse\_super} parameter while enabling \texttt{resize\_inode} (i.e., `\texttt{mke2fs -O resize\_inode,\^{}sparse\_super}').
   
By generating \textit{invalid} configuration states on purpose, we enable  examining the (mis)configuration handling of the target system. Note that this policy is inspired by the previous work on simulating human errors in configuration~\cite{Conferr-DSN08}. However, different from the relatively shallow violations (e.g., typos) which have been largely handled in matured systems, we consider more subtle violations that involve non-trivial dependencies.  

\smallskip
In addition, to provide more flexibility for different use cases, the \textit{State Generator} supports customizing the two basic policies further with different tradeoffs (e.g.,  the number of parameters  to consider, the type of dependency (i.e., SD/CPD/CCD) to use).  
As mentioned, a key challenge with analyzing configurations  of file systems is that the space is too huge to exhaust. For example, \texttt{mke2fs} itself has more than 8 trillion possible parameter combinations. With the dependency guidance, \textsc{ConfD} can reduce the space to hundreds or tens of thousands depending on the use case (\S\ref{sec:plugins}), which makes the configuration testing much more manageable in practice. And as will be shown in \S\ref{sec:dependency-aware-vs-agnostic}, the dependency-guided state generation will be more effective than dependency-agnostic alternatives for exposing configuration issues.

\subsubsection{{User Input}}
\label{sec:annotation}

\textit{ConfD-core} needs three types of input information 
from the user, which can be specified in one single JSON file.
First, to start the taint analysis, the \textit{Taint Analyzer} needs a function name as the entry point. In the case of a utility program, the function (which may invoke sub-functions) is expected to be the major function for processing configurations. In the case of the file system itself, the function can be either a function for processing configurations, or a function that is interesting (e.g., a newly added FS function). 

Second, the taint analysis also requires the names of the variables representing the configurations and the superblock in the source code, which are often different across programs based on our experience on Ext4 and XFS ecosystems. Third, to generate valid configuration states, the \textit{State Generator} needs the command-line syntax of FS configurations. Note that all the input  can be specified in the JSON format, and it is a one-time effort for each program to be analysed.

\subsection{Leveraging Configuration Dependencies}
\label{sec:leverage}    
\label{sec:plugins}

\begin{table*}[t]
	\small
	\begin{center}
		\begin{tabular}{ c | c | c | l  }
 	\textbf{Plugin ID}  & \textbf{Description}  & \textbf{Base Tool (type)}  & \textbf{\textsc{ConfD} Plugin} \\
 	\hline
 	\#1   & Configuration specification checker for Linux file systems   & N/A &  \texttt{ConfD-specCk} \\
 				\hline
	\#2     & Misconfiguration handling checker for Linux file systems     & N/A & \texttt{ConfD-handlingCk}\\
				\hline
			\#3    & An open-source fault injector for file system  utilities & 
 \texttt{rfsck}~\cite{OmFAST18} (R) & \texttt{ConfD-rfsck}   \\
          \hline
\#4   & An open-source fuzzer for file systems & \texttt{gt-hydra}~\cite{hydra-2019} (R)  & \texttt{ConfD-gt-hydra}  \\

	\hline
	
 \#5  &  Regression test suite for Linux file systems (including Ext4, XFS)  & \texttt{xfstests}~\cite{xfstest} (S)  & \texttt{ConfD-xfstests}\\
			\hline
\#6    & Regression test suite for Ext4 utilities & \texttt{e2fsprogs/tests}~\cite{e2fsprogs-test} (S)  & \texttt{ConfD-e2fsprogs}  \\
			\hline
\#7    & Regression test suite for ZFS file system & \texttt{zfs/tests}~\cite{zfs} (S)  & \texttt{ConfD-zfstests}  \\
                \hline
		\end{tabular}
	\end{center}
	\caption{ {\bf Summary of \textsc{ConfD} Plugins.} `Base Tool'  means existing tools that have been integrated with \textsc{ConfD} through the corresponding plugins; `R' means open-source Research prototype, `S' means Standard test suites for file systems and utilities.
}
	\label{tab:plugins}
\end{table*}

The dependency information and the dependency-guided configuration states may be used in different ways  to address different issues~\cite{Conferr-DSN08,spex,cdep}.  
As mentioned in \S\ref{sec:related},    there are existing efforts to improve FS ecosystems which cover a wide range of techniques including fault injection~\cite{iron05,OmFAST18}, fuzzing~\cite{xu2019fuzzing,hydra-2019}, regression test suites~\cite{xfstest,e2fsprogs-test},  etc.
While these tools are excellent for their original design goals, they are mostly agnostic to configuration dependencies and thus cannot address tricky configuration-related issues  effectively. 
The \textsc{ConfD} plugin interface is designed to bridge the gap by introducing dependency awareness to the traditional methodologies and thus amplify the effectiveness.

The current prototype of \prj includes seven plugins for file systems.
As   summarized in Table~\ref{tab:plugins},  the first two plugins (\#1 and \#2) are built from scratch, the next two plugins  (\#3 and \#4) are based  on open-source research prototypes (R), and the last three (\#5, \#6 and \#7) are designed for enhancing standard regression test suites (S). We discuss them in more details below:  

\smallskip
\noindent
\textbf{Plugin \#1: Configuration Specification Checker.}
The specifications for the configurations of Linux file systems are maintained through the Linux man-pages project~\cite{linuxmanpage}.
Unfortunately, due to a variety of reasons (e.g., constant system upgrades, feature additions, bug fixes),  the specifications may become inaccurate easily,  which may confuse end users and/or lead to configuration-induced failures~\cite{ctest-OSDI20,maintaininglinuxmanpage}.
The \texttt{ConfD-specCk} plugin is designed to mitigate the problem. It parses the Linux man-pages related to the file system configurations (e.g., \texttt{mke2fs}, \texttt{mkfs.xfs}) and checks a subset of multilevel dependencies (Table~\ref{tab:dependencies}) based on keywords. For example, \texttt{resize\_inode} and \texttt{meta\_bg} cannot be enabled together for \texttt{mke2fs} (i.e., CPD),  so  \texttt{meta\_bg} should appear in the description of \texttt{resize\_inode} with `disable' (or similar keywords) and vice versa. Similarly,   value ranges (i.e., SD) and other value dependencies (e.g., \texttt{cluster\_size} needs to be `equal' or `greater' than \texttt{block\_size}) should also be specified in the descriptions accordingly. Such  dependencies from man-pages are stored in the JSON format for further comparison with the dependencies extracted from the source code by \textit{ConfD-core} (\S\ref{sec:extraction}). A mismatch implies a potential specification issue.

\smallskip
\noindent
\textbf{Plugin \#2: Misconfiguration Handling Checker.}
A well designed  file system should be able to handle wrong configurations  from end users (either by mistake or by intention) gracefully.
Failing to handle misconfigurations elegantly implies \textit{misconfiguration vulnerabilities}  that could hurt system reliability and/or security~\cite{spex}. 
The  \texttt{ConfD-handlingCk} plugin is designed to expose the potential issues in misconfiguration handling. 
Thanks to the built-in `Violating Dependency' policy (\S\ref{sec:stategenerator}),  the plugin can directly leverage the {invalid} configuration states generated by   \prj which violate inherent configuration dependencies. It applies such automatically generated misconfigurations to drive the target file systems and utilities, and records  the symptoms accordingly for post-moterm analysis. 

\smallskip
\noindent
\textbf{Plugin \#3: Dependency-aware Fault Injector.}
Fault injection techniques have  been applied to improve both file systems and utilities~\cite{iron05,OmFAST18,Thanu-OSDI14-Crash,ShehbazJaffer-ATC19-StudyFSonSSD,e2fsprogs-test}. 
By systematically generating corrupted file system states, they enable analyzing the robustness of FS ecosystems thoroughly. 
However, given the complexity of file system metadata, 
one open challenge is how to generate vulnerable states efficiently.

To mitigate the challenge,
we integrate one open-source fault injector \texttt{rfsck}~\cite{OmFAST18} with \textsc{ConfD} through the \texttt{ConfD-rfsck} plugin.
Instead of relying on the default configuration, \texttt{ConfD-rfsck} leverages  dependency-guided configurations  to generate  input  images to initiate the fault injection campaign. Since the input images are configured with dependent parameters identified by \textsc{ConfD}, they represent more complicated states that are more difficult to remain consistent under fault. 
Note that the plugin only needs to provide an FS image with a different configuration as input to \texttt{rfsck}. No modification to the source code of  \texttt{rfsck} is required. 
As will be shown in~\S\ref{sec:results-issues}, this simple strategy can help trigger vulnerabilities  effectively.

\smallskip
\noindent
\textbf{Plugin \#4: Dependency-aware FS Fuzzer.}
Fuzzing techniques have also been applied to improve the reliability of file systems~\cite{Janus,hydra-2019}. 
Nevertheless, fuzzing file systems is still challenging due to  
the lengthy  state exploration time needed to exercise a practical file system under each configuration  (e.g., it may take multiple weeks to trigger one  bug~\cite{hydra-2019}). 
In other words,
the time penalty for exploring a less interesting configuration state is high.

To mitigate  the challenge,
we integrate one open-source fuzzer \texttt{gt-hydra}~\footnote{To avoid confusion, we use \texttt{gt-hydra} to refer to the Hydra fuzzing framework created by GaTech researchers~\cite{hydra-2019}, and use  FB-HYDRA to refer to the Hydra configuration management framework created by Facebook~\cite{Yadan2019Hydra}.}
with \prj through the \texttt{ConfD-gt-hydra} plugin.   
Similar to 
plugin \#3,
\texttt{ConfD-gt-hydra} leverages dependency-guided configurations generated by \textsc{ConfD} to create  FS images with more complicated dependencies and thus more chances of vulnerability for fuzzing. 
The plugin only changes the configurations of the input images for \texttt{gt-hydra}; no modification to the source code of the base tool is needed.

\smallskip
\noindent
\textbf{Plugin \#5, \#6, \#7: Dependency-aware Regression Test Suites}.
Besides research prototypes, there are  standard regression test suites developed for file systems (e.g., \texttt{xfstests}~\cite{xfstest} and \texttt{e2fsprogs/tests}~\cite{e2fsprogs-test}), which include carefully designed workloads and test oracles to ensure the quality of the target.
Nevertheless, existing test suites 
only use a subset of configuration parameters  and they are mostly  dependency-agnostic.
To address the limitation,  we first create two plugins (\#5, \#6) for enhancing the testing of Linux file systems and utilities:  \texttt{ConfD-xfstests} and \texttt{ConfD-e2fsprogs}, for  \texttt{xfstests} and \texttt{e2fsprogs/tests} respectively.  
The plugins  scan the test scripts and {automatically} replace the built-in FS configurations of the test cases with the configuration states generated by \textsc{ConfD}. The two plugins use the `Follow Dependency' policy of  \textsc{ConfD}  to drive the test cases deeply into the target functionalities without early termination due to superficial configuration errors.  In doing so, we reuse the well designed test logic and enhance the test suites with dependency awareness. 
If any test case fails with the \textit{valid}  configurations provided by \textsc{ConfD}, the result is saved for postmortem analysis. 
 
In addition to Linux file systems, there are other popular file systems which include sophisticated features and configurations. One representative example is ZFS~\cite{zfs}, which was originally built on Solaris and has been ported to many Unix-like systems. To explore the extensibility of  \textsc{ConfD}, we create  the plugin \#7 for ZFS based on the ZFS test suite. Different from  traditional Linux file systems which include a unified test suite (i.e., \texttt{xfstest} for Ext4, XFS, etc.), ZFS includes a dedicated test suite for testing  ZFS only (i.e.,  \texttt{zfs/test}~\cite{zfs}). The ZFS test suite consists of test cases for both ZFS file system and ZFS utilities, and the testing logic  may cover both functional correctness and system performance. 
Since our focus is to identify configuration related correctness issues, we only consider functionality testing in this paper, and leave performance testing  as future work.
To test ZFS with dependency aware configuration states, we introduced plugin \#7 ConfD-zfstests. 
Similar to plugins \#5 and \#6, this ZFS plugin  scans the testing script of ZFS test suite and automatically replaces the existing configuration state with a dependency-aware configuration state for testing. The testing output, including any error logs, is saved in a separate directory for further investigation. 
 
\smallskip    
Note that \textsc{ConfD} plugins are not limited to the seven examples above. By modularizing the core module of  \textsc{ConfD} (Figure~\ref{fig:overview}), we expect that other software may benefit from  \textsc{ConfD} conveniently via plugins (see \S\ref{sec:discussion} for more discussion). \\

\subsection{Optimization for Reducing False Positive Rate of Dependency-Aware Regression Testing}
\label{sec:reducefalsepositives}

In our early experimentation~\cite{mahmud2023confd} with \textsc{ConfD}-enhanced  regression test suites for Linux file systems (i.e.,  \texttt{ConfD-xfstests} and \texttt{ConfD-e2fsprogs}), we observed that the test cases may fail  due to various reasons including timing at \texttt{mount}, bugs in FS source code, incompatible environmental variables, etc.
Besides these typical reasons, 
the dependency awareness introduced by \textsc{ConfD} 
adds another potential source of failure: 
as  the plugins need to modify the source code of test cases to change the  built-in configurations (\S\ref{sec:plugins}), the modification itself might cause unexpected failures. 
We found that this was mainly caused by the irregularity of syntax and function invocations used in the test suites, which makes it difficult to change existing configuration parameters automatically and perfectly without breaking the test cases. For clarity, we call this type of undesirable failures as \textit{plugin-caused failures}.

The \textit{plugin-caused failures} may contribute to a non-negligible percentage of  false positives of regression testing (e.g., 30.43\% for Ext4 based on our measurement). 
One  challenge with identifying configuration issues using  regression test suites is the large number of test results that need manual  analysis and validation with strong domain knowledge.
This is exacerbated by the fact that running a complete set of vanilla tests  with individual configurations (e.g., default ones) is time-consuming (e.g., it may take hours or even a full day for Ext4, depending on the compute resources available\cite{storagetesting}). In other words, \textit{plugin-caused failures}  may lead to substantial waste of time and resources for regression testing of file systems~\cite{howtofindExt4bugs}, and thus they should be avoided as much as possible.

To reduce \textit{plugin-caused failures}, we analyzed the source code of the regression test suites in depth.
In terms of \texttt{ConfD-xfstests} (which includes test cases for both Ext4 and XFS), we identified three common patterns for causing \textit{plugin-caused failures}:
(1) The original test relied on a specific configuration parameter which is not presented in the dependency-aware configuration state; (2) The original test invoked a specialized function expecting a specific size parameter; and (3) corner cases of  syntax which were not recognized by \textsc{ConfD}. Based on these observations, we optimized the  plugin to handle tests with existing parameters  by adding a new mode: besides replacing existing parameters with dependency-aware configurations, the plugin may generate a new test case by  appending non-conflicting configurations (based on the dependency information extracted by \prj) to the existing parameters. 
In addition, we also enhanced \texttt{ConfD-xfstests}  to handle additional  functions and special syntax. As we will show in \S\ref{sec:FalsePositiveExp}, these simple optimizations can reduce the false positive rate of \texttt{ConfD-xfstests}  effectively.

In terms of \texttt{ConfD-e2fsprogs},  the test cases are grouped based on the feature being tested. For example, the "f" series test the \texttt{e2fsck} utility of Ext4 for various \texttt{mkfs}
features and corruption scenarios. We found that 8.7\% of the tests failed after adding dependency awareness, but none of 
them belonged to {plugin-caused failures}.   
Therefore,  no further optimization is needed to reduce the false positive rate of  \texttt{ConfD-e2fsprogs}.

\subsection{Extending \prj for Other Storage Systems}
\label{sec:DBextension}
Besides file systems, there are many other data management systems (e.g., databases) which also contain a rich set of tunable features and configuration parameters. To explore   configuration dependency in the broader context beyond file systems, we have analyzed the configurations of  WiredTiger~\cite{wt}, the default NoSQL storage engine for MongoDB~\cite{mongodb}. Similar to file systems,  WiredTiger stores a set of initial configuration parameters along with other  metadata  in a dedicated file. 
Different from file systems, WiredTiger maintains much  dependency information in a dedicated array data structure. Configuration parameters specified by the end users  as input may be validated against the dependency information stored in the  array before creating a database instance. Moreover, the dependencies may be extracted and used at runtime (e.g., opening database, creating a table or transaction). 
Based on this  observation, we extended \prj-core  to parse  the configuration array and extract the self and cross-parameter dependencies automatically. Note that due to the complex usage of configuration parameters throughout the source code of WiredTiger, our current extension  only works for a limited subset of configurations related to database creation. We leave the  extraction  of other database configurations  (and the corresponding plugin)  as future work.

\section{Experimental Results}
\label{sec:results}

In this section, we describe the experimental results of applying \textsc{ConfD} to analyze Ext4, XFS, ZFS and WiredTiger.
First (\S\ref{sec:results-dependencies}), we show that \textsc{ConfD} can extract 160 multilevel configuration dependencies from the target systems  effectively with a low false positive rate (8.1\%).
Second (\S\ref{sec:results-issues}), we demonstrate that \textsc{ConfD} can help address configuration-related issues more effectively compared to existing dependency-agnostic solutions. 
Through the experiments, we have identified various configuration-related issues including 17 specification issues, 18 configuration handling issues, and 10 regression test failures induced by valid configurations.

 \begin{table*}[t]
	\small
	\begin{center}
		\begin{tabular}{ c | c | c | c | c | c | c | c | c}
			\textbf{Target FS } & \multicolumn{2}{c|}{\textbf{Self Dependency (SD)}} & \multicolumn{2}{c|}{\textbf{ Cross-Param. Dep. (CPD)}}  & \multicolumn{2}{c|}{\textbf{Cross-Comp. Dep. (CCD)}} & \multicolumn{2}{c}{\textbf{All Level Combined}}\\
 \cline{2-9}
		
		 \textbf{Ecosystem} & \textbf{Extracted}  &  \textbf{FP} &  \textbf{Extracted}  &   \textbf{FP}  &  \textbf{Extracted} &  \textbf{FP}   & \textbf{Extracted} &  \textbf{FP} \\
					\hline
{Ext4}  & 17 & 0  & 48 & 1 (2.1\%) & 46 & 3 (6.5\%) & 111 & 4 (3.6\%) \\
			\hline
{XFS}  & 18 & 2 (11.1\%) & 10  & 3 (30.0\%) & 15 & 4 (26.7\%) & 43 & 9 (20.9\%)\\
			\hline
{ZFS} & 4 & 0 & 1 & 0 & 0 & 0 & 6 & 0\\
                \hline
 \multicolumn{1}{r|}{\textbf{Total}} & 39 & 2  (5.7\%) & 59 & 4 (6.9\%) & 61 &  7 (11.5\%)  &  160 & 13 (8.1\%)\\
\hline
		\end{tabular}
	\end{center}
	\caption{ {\bf  Multilevel Configuration Dependencies Extracted by \textsc{ConfD}.} This table shows the numbers of multilevel  dependencies extracted from three FS ecosystems (Ext4, XFS, ZFS) automatically.  `FP' means False Positive rate.  
}
	\label{tab:accuracyofdependency}
\end{table*}

\begin{table*}[t]
	\small
	\begin{center}
		\begin{tabular}{ c | c | c }
			\textbf{Target } & \multicolumn{2}{c}{\textbf{Self Dependency (SD)}} \\
            \cline{2-3}
		      \textbf{database} & \textbf{Extracted}  &  \textbf{FP}  \\
			\hline
            {WiredTiger}  & 35 & 0   \\
			\hline
		\end{tabular}
	\end{center}
	\caption{ {\bf  Configuration Dependencies Extracted by from WiredTiger.} This table shows the numbers of configuration dependencies extracted from WiredTiger automatically.  `FP' means False Positive rate.}
	\label{tab:wiredtigerdep}
\end{table*}

\begin{table}[t]
	\small
	\begin{center}
		\begin{tabular}{ c | c | c | c | c   }
 	\textbf{Target FS }  &  \multicolumn{4}{c}{\textbf{\# of Uncorrectable Images Reported}}    \\
          \cline{2-5}
           	\textbf{ Ecosystem  }  & {\texttt{rfsck} (1) }  & \multicolumn{3}{c}{{\texttt{ConfD-rfsck} (25)}  }    \\
\hline
Ext4  & 11  & $<$ 11 (4) & $=$ 11 (4) & $>$ 11 (17)   \\
			\hline
 
		\end{tabular}
	\end{center}
	\caption{ {\bf Comparison of Two FS Fault Injectors.} 
\texttt{rfsck} explores 1 default configuration state and reports 11 uncorrectable images. \texttt{ConfD-rfsck} explores 25  configuration states;   it reports 
$>$ 11 uncorrectable images (i.e., better than \texttt{rfsck}) in 17 out of 25 configuration states.}
	\label{tab:result-rfsck}
\end{table}

\subsection{Can \textsc{ConfD} extract multilevel dependencies?}
\label{sec:results-dependencies}

Table~\ref{tab:accuracyofdependency} summarizes the multilevel configuration dependencies extracted by \textsc{ConfD} from Ext4, XFS and ZFS   {automatically}. 
As shown in the table, we were able to extract 160  unique  dependencies in total, including 39 Self Dependency (SD), 59 Cross-Parameter Dependency (CPD), and 61 Cross-Component Dependency (CCD). 
The multilevel dependencies have been observed on  both Ext4 and XFS,  which is consistent with our manual study (\S\ref{sec:bugstudy}).  
 
We manually examined all the 160 dependencies extracted by \prj automatically and found that the overall false positive rate is 8.1\% (13/160), which is similar to that of the previous work on analyzing configuration constraints in other software  systems~\cite{cdep,spex}. Note that \prj is designed to handle the  unique configuration methods of FS ecosystems (\S\ref{sec:background} and \S\ref{sec:findings})  which is arguably more challenging to analyze  compared to the targets of existing work.

 Table~\ref{tab:wiredtigerdep} summarizes the extracted dependencies from WiredTiger~\cite{wt}. We were able to extract 35 self dependencies from the source code and  verified them to be correct without any false positives. This result  is expected because we focus on the database creation subsystem of WiredTiger (\S\ref{sec:DBextension}).  Broadening the scope to cover other subsystems will likely extract   more dependencies including  cross-component dependencies, which we leave as future work.
  
\vspace{-0.1in}
\subsection{Can \prj help address configuration issues?}
\label{sec:results-issues}
\subsubsection{Dependency-agnostic vs. Dependency-guided}
\label{sec:dependency-aware-vs-agnostic}

In this section, we compare the effectiveness of two open-source research prototypes (i.e.,  \texttt{rfsck}~\cite{OmFAST18} and \texttt{gt-hydra}~\cite{hydra-2019}) with and without  \prj support. We focus on the two research prototypes and the corresponding plugins for   comparison   because they provide quantitative metrics to measure the effectiveness straightforwardly. We defer the results of other plugins to the next section. 

\begin{table}[t]
	\small
	\begin{center}
		\begin{tabular}{ c| c | c | c    }
 \textbf{ID} &	\textbf{Symptom of  }  &  \multicolumn{2}{c}{\textbf{Triggered?}}    \\
          \cline{3-4}
          & 	\textbf{Uncorrectable Corruption   }  & {\texttt{\space \space rfsck \space \space}}  & {{\texttt{ConfD-rfsck}}  }    \\
\hline
1  & Unable to mount the FS & N  &  Y  (6) \\
			\hline
2  & Invalid file data  & N  &  Y  (24) \\
			\hline
3  & Truncated file data  & Y (11) &  Y (250)   \\
			\hline
\multicolumn{2}{r|}{\textbf{Total}} & 11  &  280    \\
			\hline
		\end{tabular}
	\end{center}
	\caption{ {\bf Comparison of Corruption Symptoms Triggered.} 
\texttt{ConfD-rfsck} triggered (`Y') more types of corruptions. The counts are in parentheses.
}
	\label{tab:result-rfsck-symptom}
\end{table}

In the first experiment, we applied fault injectors \texttt{rfsck}  and \texttt{ConfD-rfsck} to analyze Ext4 and its checker utility \texttt{e2fsck}. The fault injectors  interrupt the checker operation and examine if the interrupted checker   could lead to uncorrectable corruptions on the file system (i.e., cannot be fixed by another run of checker). They report the number of repaired FS images containing uncorrectable corruptions (i.e., ``uncorrectable image''). Each uncorrectable image implies a vulnerability in the FS ecosystem that could lead to data loss~\cite{OmFAST18}.  

The result of the experiment is summarized in Table~\ref{tab:result-rfsck}. \texttt{rfsck} reports 11 uncorrectable images with the default configuration.  \texttt{ConfD-rfsck} can explore different configuration states and we analyze the reports generated under 25 configuration states for comparison. In 4   out of the 25 states, \texttt{ConfD-rfsck} generates less than 11 uncorrectable images; in   4 states,  \texttt{ConfD-rfsck} generates the same amount of  uncorrectable images (i.e., `$=$ 11'); 
 in the majority states (17), \texttt{ConfD-rfsck}  generates more uncorrectable images (i.e., `$>$ 11'), which suggests  it is more effective in exposing potential vulnerabilities in the FS ecosystem. 
Table~\ref{tab:result-rfsck-symptom} further compares the symptoms of uncorrectable corruptions triggered by \texttt{rfsck}  and \texttt{ConfD-rfsck}. Overall, \texttt{ConfD-rfsck} triggers three different types of symptoms,  while \texttt{rfsck} only triggers one symptom in our experiment.  
 Since different symptoms typically imply different vulnerabilities in metadata protection and/or recovery in the FS ecosystem, the result also suggests that   the dependency-guided configuration states used by  \texttt{ConfD-rfsck} can help improve the effectiveness of \texttt{rfsck}.

\begin{table}[t]
	\small
	\begin{center}
		\begin{tabular}{ c | c | c    }
 	\textbf{Target FS }  &  \multicolumn{2}{c}{\textbf{\# of Issues Reported (in two weeks)}}    \\
          \cline{2-3}
           	\textbf{   }  & {\texttt{\space \space \space gt-hydra \space \space \space}}  & {\texttt{ConfD-gt-hydra}}    \\
\hline
Ext4  & 1  &  17      \\
			\hline
		\end{tabular}
	\end{center}
	\caption{ {\bf Comparison of Two FS Fuzzers.} 
\texttt{ConfD-gt-hydra} reports more hangs given the same fuzzing time.
}
	\label{tab:result-gt-hydra}
\end{table}

In the  second experiment, we applied \texttt{gt-hydra}  and \texttt{ConfD-gt-hydra} to fuzz the Ext4 file system.  
The fuzzers systematically generate various inputs (i.e., FS metadata corruptions and system calls) to explore different code paths in the file system for triggering latent bugs~\cite{hydra-2019}.
We run each  fuzzer continuously for two weeks.  The fuzzers report the number of reliability issues  detected on the target file system within the running period. The issues may include different types depending on the bug checkers used. We use the default \textsc{SymC3} checker which can detect crash inconsistency bugs. Meanwhile, since  the fuzzers are based on the AFL fuzzer ~\cite{afl}, they also report crash and hang issues (detected by AFL) by default. 
Note that the only difference \texttt{ConfD-gt-hydra} introduces is the dependency-guided configurations,
i.e., it does not change the test logic or criteria for reporting issues. Therefore, both the types of issues (e.g., `crash', `hang', `crash inconsistency') and the number of issues reported can be 
used as the metric to evaluate effectiveness.

The result of the fuzzing experiment is summarized in Table~\ref{tab:result-gt-hydra}.
To make the comparison fair, we limit the two fuzzers to the same total execution time (i.e., two weeks each).
We set the \texttt{ConfD-gt-hydra} to switch to a new dependency-guided configuration state every 12 hours, which leads to 28 critical configuration states being explored within two weeks. While each configuration in \texttt{ConfD-gt-hydra} is  explored with only 1/28 of the  time used by \texttt{gt-hydra} for its configuration, the overall result of \texttt{ConfD-gt-hydra} is   better:  \texttt{gt-hydra} only detects 1 issue on Ext4 by the end of the two week period, while \texttt{ConfD-gt-hydra} detects 17 issues in total. 
Interestingly, all issues   reported in the experiment are `hang'. 
This is expected because triggering  more complicated semantic bugs may require multiple weeks. 
 
In summary, the two sets of comparison experiments  above show that  \prj can  amplify the effectiveness of existing FS tools for identifying vulnerabilities quickly, which is particularly valuable for time-consuming methodologies like fault injection or fuzzing. 
Note that in all experiments, we do not randomly generate combinations of configurations. This is because a naive algorithm without any knowledge of inherent dependencies can easily lead to time-wasting configurations, as   will be  demonstrated further  in \S\ref{sec:state-generation-vs-fb-hydra}.

\subsubsection{Summary of Configuration Issues}
\label{sec:summary-issues}

\begin{table}[t]
	\small
	\begin{center}
		\begin{tabular}{ l   | c | c | c }
		 	\textbf{\textsc{ConfD} Plugin}     &\multicolumn{3}{c}{\textbf{\# of Issue Reported}}  \\
		 	  \cline{2-4}
   \textbf{(Type of  Issue Reported)}   & \textbf{Ext4}  &  \textbf{XFS}  & \textbf{Total} \\
          \hline
 \texttt{ConfD-specCk}    (undoc./wrong dep.)  &    13  & 4  &   17 \\
			\hline
\texttt{ConfD-handlingCk}  (bad reaction)   & 13  &  5   &  18\\
          \hline
 \texttt{ConfD-xfstests}   (test case failure)   &    5  & 4  &   9 \\
           \hline
 \texttt{ConfD-e2fsprogs}   (test case failure)  &    1  & N/A  &   1 \\
           \hline
 \texttt{ConfD-rfsck}   (uncorrectable image)    &    280  & --  &   280 \\
           \hline
 \texttt{ConfD-gt-hydra}  (hang)   &    17  & --  &   17 \\
  	\hline
		\end{tabular}
	\end{center}
	\caption{ {\bf Summary of Issues.} 
This table summarizes configuration-related issues observed via  \prj plugins. 
}
	\label{tab:result-summary}

\end{table}

\begin{table}[t]
	\small
	\begin{center}
		\begin{tabular}{ c | c | c | c | c }
		 	\textbf{Target FS}  &\multicolumn{3}{c|}{\textbf{\# of Undocumented/Wrong Dep.}} & \textbf{Total} \\
		 	  \cline{2-4}
 	\textbf{Ecosystem}   & \textbf{\space \space \space \space SD \space \space \space \space}  & \textbf{\space \space \space \space CPD \space \space \space \space}  & \textbf{CCD}  & \\
          \hline
   Ext4  &  7  &    4  & 2  &   13 \\
			\hline
XFS    & 2 &  2 &  0   & 4 \\
	\hline
  \multicolumn{1}{r|}{\textbf{Total}} & 9 & 6  &     2& 17 \\
  	\hline
		\end{tabular}
	\end{center}
	\caption{ {\bf Specification Issues.} 
This table summarizes the undocumented or wrong dependencies observed. `SD', `CPD', and `CCD' are   defined in Table~\ref{tab:dependencies}. 
}
	\label{tab:documentation-issue}
\end{table}

\begin{table*}[t]
	\small
	\begin{center}
		\begin{tabular}{ c | c | c | c  }
 	\textbf{ID}  & \textbf{Reaction}  & \textbf{Description} & \textbf{Observed?}     \\
          \hline
 1   & Early Termination   & the  utility program exits w/o pinpointing the configuration error &  Y \\
			\hline
		 2    & Functional Failure  & the utility fails functional testing w/o pinpointing the configuration error &   Y \\
	\hline
  3  &  Silent Violation & the system changes input configurations to different values w/o notifying users & Y \\
			\hline
 4    &  Silent Ignorance  & the system  ignores input configurations & N   \\
			\hline
  5    & Crash/Hang  & the system crashes or hangs  & N   \\
			\hline		
6    &  Partial Report  & the utility  partially identify the violated configuration dependencies  & Y    \\
			\hline
		\end{tabular}
	\end{center}

	\caption{ {\bf Suboptimal Reaction of Configuration Dependency Violation.}  This table summarizes the bad handling behaviors observed when the  configuration dependencies are violated. The first five   are based on the definitions from    \cite{spex}.  
}
	\label{tab:handling}
\end{table*}

Table~\ref{tab:result-summary} summarizes 
the configuration-related issues triggered by \prj plugins in our experiments. Overall, we  observed more than 300   issues of various types.
The issues are diverse because the plugins are created for different purposes  or based on different base tools  (Table~\ref{tab:plugins}). 
Note that all the issues require   dependency-guided configuration states generated by \prj to manifest. In other words, continuously running the original research prototypes or standard test suites cannot expose the issues.
Also, since we do not change  the test logic of the base tools,  the enhancement  is purely contributed by the dependency information from \prj. Since   \texttt{ConfD-rfsck} and  \texttt{ConfD-gt-hydra} have been discussed in \S\ref{sec:dependency-aware-vs-agnostic}, we focus on others  below.

Table~\ref{tab:documentation-issue} summarizes the specification issues detected by  \texttt{ConfD-specCk}.
We have identified 17 inaccurate specification issues in total.
The issues mainly manifest as undocumented critical dependencies or wrong dependencies, which may occur to both Ext4 and XFS and involve SD, CPD, and CCD. For example, there is a CPD extracted by \prj which specifies that two  parameters of   \texttt{mke2fs} (i.e., \texttt{meta\_bg} and \texttt{resize\_inode})  cannot be used together, but this CPD is missing from the Linux man-pages.
 As another example, there is a CCD which implies that \texttt{resize2fs} may not be used for Ext4 when the \texttt{bigalloc} feature is enabled through \texttt{mke2fs}. Violating the CCD may corrupt the file system, which is unfortunately not mentioned in the specification. 

Table~\ref{tab:handling} summarizes the suboptimal handling of misconfigurations identified through \texttt{ConfD-handlingCk}. 
We follow the criteria in the literature~\cite{spex}: when a misconfiguration occurs (i.e., a dependency is violated),  the system should pinpoint either the offending parameter's name/value or its location information; failing to do so implies misconfiguration vulnerabilities. 
Specifically, there are six types of misconfiguration vulnerabilities based on different reactions, including  
`Early Termination', `Functional Failure', `Silent Violation',  `Silent Ignorance', `Crash/Hang', and `Partial Report'. The first five types are based on the definitions from~\cite{spex}, while the last one is unique in our study because we consider more complicated multilevel dependencies.   

As an example, the \texttt{mke2fs} parameter \texttt{-E encoding}  enables the \texttt{casefold} feature and set the encoding in Ext4. But if the user tries to disable the \texttt{casefold} feature when using the \texttt{-E encoding}, instead of showing an error or warning, the utility enables the \texttt{casefold} feature silently without informing the user. We consider this as  `Silent Violation'.

When more than one dependency is violated, utilities often only show a partial message (i.e., `Partial Report'). 
For example, the \texttt{mkfs.xfs} parameter \texttt{sunit}  involves two dependencies:
(1) it does not allow unit suffixes, and (2) it cannot be specified together with \texttt{su}. But when both dependencies are violated, the utility may only show one of the violations.

 In total, we have observed 4 out of the 6 types of suboptimal reactions, which suggests that FS ecosystems are not immune  from misconfiguration vulnerabilities reported in other practical systems. Note that \texttt{ConfD-handlingCk} leverages the static analysis of \textsc{ConfD} to violate specific dependencies carefully, which avoids many duplicate and valid configuration states for testing. This reduces the manual effort needed for the post-mortem analysis.    

In terms of \texttt{ConfD-xfstests} and \texttt{ConfD-e2fsprogs}, we have observed 10 new test case failures (9 from \texttt{ConfD-xfstests} and 1 from \texttt{ConfD-e2fsprogs}) which can be induced by \textit{valid} configuration states generated by \prj.
For example,  
\texttt{ConfD-xfstests} triggers an Ext4  corruption when applying the online defragmentation tool \texttt{e4defrag} to the file system with  the \texttt{bigalloc} feature enabled.
Note that a FS test case may involve multiple utilities. 
Due to the complexity of the test case and the FS ecosystems, a test case may fail for various subtle reasons (e.g., timing at \texttt{mount}) in practice, which is time-consuming to diagnose even for developers~\cite{howtofindExt4bugs} as discussed in \S\ref{sec:reducefalsepositives}. In our experiments, we observed more than 10 newly failed test cases after changing with valid configurations. 
We only count the cases that we have manually verified and reproduced at the time of this writing.
Also, since  \prj  limits the change to the configuration states without modifying the test logic, it may help narrow  down the root cause of a test case failure to the configuration-related code paths.

Table ~\ref{tab:testcaseresponse} further breaks down the result of test cases for different target software during  regression testing experiments.  \texttt{xfstests} classifies test cases into different groups based on different target software. For Ext4 and XFS, it has 46 and 97 test cases respectively. The Generic group has a total of 181 test cases which are designed to test all the file systems in the test suite. 
As shown in the last column of the table, \texttt{ConfD-xfstests} has triggered  failures induced by \textit{valid} configurations  in all three test case groups (i.e., 4 for Ext4, 4 for XFS, and 1 for the Generic group).

\begin{table*}[t]
	\small
	\begin{center}
		\begin{tabular}{ c | c | c | c | c }
& \textbf{Target FS} & \multicolumn{2}{c|}{\textbf{Testcases \#}} &  \textbf{Dep. Aware} \\
        \cline{3-4}
    \textbf{Plugin} & \textbf{Ecosystem}  & \textbf{\space \space \space \space \space Total \space \space \space \space \space}  & \textbf{\space \space \space \space \space Used \space \space \space \space \space} & \textbf{Config. States \#}  \\
          \hline
ConfD-zfstests & ZFS  & 1581   & 154 & 167 \\
	\hline
		\end{tabular}
	\end{center}
	\caption{ {\bf Summary of ConfD-zfstests.}  This table summarizes the statistics of Plugin \#7 ConfD-zfstests.}
	\label{tab:zfstests}
\end{table*}

For \texttt{ConfD-zfstests}, we have not observed any test case failure. The reason behind could be the less number of configuration dependencies we could extract from the source code since the effectiveness of the plugins are dependent on the critical configuration dependencies extracted from the source code. 
Table ~\ref{tab:zfstests} shows the statistics of running Plugin \#7. We have tried 167 dependency-aware configuration states based on the configuration dependency extracted by ConfD-core on 154 test cases from ZFS test suite. Since our focus was \texttt{Zfs-create} configuration parameters and ZFS doesn't group test cases for different utilities, we separated the test cases containing \texttt{Zfs-create} command from the ZFS test suite and only ran our configuration states on the selected ones. ZFS test suite has a total of 1581 test cases and separating the test cases bases on our interest saves time spent running irrelevant test cases and doesn't incur any false positives. Additionally, our plugin also allows to adding new test cases same as the original test suite.

\begin{table*}[t]
	\small
	\begin{center}
		\begin{tabular}{ c | c | c | c |  c }
        \textbf{Plugin}  & \textbf{ Target Software}  & \textbf{\space \space Total Test \# \space \space} & \textbf{\space \space Failed Test \space \space}  & \textbf{ Induced by Config.}     \\
          \hline
 ConfD-xfstests   & Ext4   & 46 &  18   & 4  \\
			\cline{2-5}
		     & XFS  & 97 &   11  & 4   \\
             \cline{2-5}
             & Generic & 181 & 173 &   1   \\
	\hline
   ConfD-e2fsprogs   & e2fsck   & 23 & 2  & 0  \\
			\cline{2-5}
		     & resize2fs  & 19 &   6  & 1  \\
			\hline
		\end{tabular}
	\end{center}

	\caption{ {\bf Test Case Result for Different Target Software.} 
    This table breaks down the result of regression tests for different target software.
    The last column shows the number of failures induced by valid configurations after our manual validation. 
}
	\label{tab:testcaseresponse}
\end{table*}

\begin{table*}[t]
	\small
	\begin{center}
		\begin{tabular}{ c | c | c | c | c }
         & & & \multicolumn{2}{c}{\textbf{False Positives}} \\
                    \cline{4-5}
 	\textbf{Plugin}  & \textbf{Target}  & \textbf{Total} & \textbf{Before } & \textbf{After}     \\
        \textbf{ }  & \textbf{Software}  & \textbf{test cases \#} & \textbf{Optimization} & \textbf{Optimization}     \\
          \hline
ConfD-xfstests   & Ext4   & 46 & 14 (30.43\%) & 1 (2.17\%)\\
			\cline{2-5}
 & XFS  & 97 &  7 (7.22\%) & 4 (4.12\%) \\
	\hline
		\end{tabular}
	\end{center}

	\caption{ {\bf Summary of False Positive Reduction.}  This table summarizes the number of false positives before and after our optimization to the testing script. 
}
	\label{tab:FPreduction}
\end{table*}

\vspace{-0.05in}
\subsubsection{Reduction of False Positive Rate.}
\label{sec:FalsePositiveExp}
As discussed in \S\ref{sec:reducefalsepositives}, it is important to avoid \textit{plugin-caused failures} to reduce false positives for regression testing. 
 Table ~\ref{tab:FPreduction} summarizes the false positive rate before and after our optimization for \texttt{ConfD-xfstests} (\S\ref{sec:reducefalsepositives}). 
Before the optimization,  
30.43\% (14/46) of the Ext4, and 7.22\% (7/97) of the XFS tests exhibited plugin-caused failures.
 After the optimization, we were able to reduce the percentages to 2.17\% (3/46) and 4.12\% (6/97) for Ext4 and XFS respectively,  which results in improvement of false positive rate by 92.87\% (for Ext4) and 42.93\% (for XFS) respectively. 

An example of the false positive reduction,  we examines a case from \texttt{ext4/011} in details. The original test invokes \texttt{mkfs} with two necessary configurations (\texttt{-O mmp} and \texttt{-E mmp\_update\_interval=2}). Before the optimization, these configuration parameters were blindly discarded and  replaced with the configurations generated by \prj, which leads to a plugin-caused failure. After the optimization, \texttt{ConfD-xfstests} maintains these existing parameters to satisfy the necessary condition and append additional parameters of interest that do not conflict with the original test, which effectively eliminates the false positive instance.

\vspace{-0.05in}
\subsubsection{State Generation:  FB-HYDRA vs. \textsc{ConfD}}
\label{sec:state-generation-vs-fb-hydra}
\vspace{-0.05in}

\begin{table}[t]
	\small
	\begin{center}
		\begin{tabular}{ c | c | c | c  }
 	\textbf{Framework}  & \textbf{\space \space \# of States \space \space}  & \textbf{\# of Duplicate}  &  \textbf{\# of Invalid} \\
          \hline
FB-HYDRA  & 56,592 &  42,745 (75.5\%)  &  15,146 (26.8\%)   \\
			\hline
		\textsc{ConfD}    & 30 &  0 &  0  \\
	\hline
 
		\end{tabular}
	\end{center}

	\caption{ {\bf Comparison of State Generation.} 
}
	\label{tab:vs-fb-hydra}
	\vspace{-0.1in}
\end{table}

One unique feature of \textsc{ConfD} is it generates configuration states based on  multilevel dependencies, which is critical for analyzing configuration issues given the huge configuration space.
To the best of our knowledge, the FB-HYDRA configuration management framework~\cite{Yadan2019Hydra} provides the most similar functionality. It includes a ``multirun'' feature to support running an application with different configurations in different runs automatically. We compare the configuration states generated by FB-HYDRA and \textsc{ConfD} in this section to demonstrate the difference.

Table~\ref{tab:vs-fb-hydra} shows  the states generated by FB-HYDRA and \textsc{ConfD} for the same program (i.e., \texttt{mke2fs}) given the same set of  configuration parameters.  For simplicity, we only use 10 parameters with limited ranges  in this experiment. 
As shown in the table, even with this simplified scenario, FB-HYDRA may generate many duplicated    or  invalid states. This is because FB-HYDRA is agnostic to the configuration constraints of \texttt{mke2fs}. 
Specifically, FB-HYDRA maintains a list for each parameter and its possible values. It  passes all lists to the \texttt{itertools.product()} function which returns the cartesian product of the values in the lists.
Such a simple algorithm is incompatible with FS ecosystems. For example, `\texttt{mke2fs -b 1024 -C 2048}' and `\texttt{mke2fs -C 2048 -b 1024}' are equivalent in practice but are considered as different in FB-HYDRA. Moreover, invalid states can easily be created by FB-HYDRA due to violation of dependencies, which suggests the importance of dependency analysis.

Note that FB-HYDRA has other features that \textsc{ConfD} does not have (e.g., Python library support). Also, FB-HYDRA supports plugins which makes it possible to benefit from the state generation of \textsc{ConfD} (see \S\ref{sec:discussion} for more discussion).
Therefore, we view FB-HYDRA and \textsc{ConfD}  as complementary.

\section{Discussions}
\label{sec:discussion}

 No study or tool is perfect, and our work is no exception. We discuss lessons learned,  limitations of \prj prototype, and a few promising extensions in this section.

\subsection{Lessons Learned}
\noindent
\textbf{Heterogeneity in storage configuration  management.}
Through this comprehensive study, we found that
although many file systems follow similar modular design principles (\S\ref{sec:background}), 
their configuration management may be implemented in various ways. Classic file systems, such as Ext4 and XFS, tend to have a relatively simpler management mechanism. For example, there is usually one main configuration function where most of the checking of configuration parameters are implemented. On the other hand, newer file and storage systems (e.g., ZFS and WireTiger) may have relatively more sophisticated  configuration management to cover a wider configuration space for more advanced features (e.g., snapshot, cloning).
For example, different from Ext4/XFS which  use \texttt{mkfs.ext4}/\texttt{mkfs.xfs} only to create the file system itself, 
ZFS supports creating various types of dataset (e.g.,  file system, volume, snapshot,  pool~\cite{zfs}) via  \texttt{zfs-create}.
All the ZFS datasets have their own sets of parameters; meanwhile, they may also be affected by a same set of shared configuration parameters.  Moreover, ZFS supports user defined configurations and inherited parameters. Instead of keeping user input values in variables (as in Ext4/XFS),  each ZFS parameter and its value is stored as a pair in a separate data structure. To manage such unique features, ZFS configuration subsystem uses various different functions to handle parameter conversion and checking, which involves both general checking for all parameters and  specific checks for a subset of special parameters.

In addition,  heterogeneity has also been observed in terms of configuration documentation. 
Such heterogeneity inevitably contributes to the complexity of configuration-oriented
testing and makes a one-size-fits-all solution almost impossible.
We hope that  our study as well as the plugin-based \prj design can help alleviate the challenge in configuration management and inspire follow up research on handling the growing heterogeneity and complexity.

\vspace{0.05in}
\noindent
\textbf{Need for better configuration design.}
An alternate perspective of the configuration challenge studied and exposed in this work is that we may have too many parameters today. One might argue that  it is perhaps better to reduce the parameters to avoid  vulnerabilities or confusions, instead of adding new configurations for more features. 
Also, one might suggest that (in theory) we can implement every utility functionality in  the file system itself 
to avoid  tricky cross-component configuration dependencies.
Essentially, these are trade-offs of the file system and configuration design that deserve more investigation from the community. We hope that by studying real-world configuration issues and releasing the \textsc{ConfD} prototype, our work can help {identify problematic configuration parameters and further help with the reduction of such parameters to} improve the configuration design  in general.\\

\vspace{-0.15in}
\subsection{Limitations}
\noindent
{\bf Limitations of the multilevel taxonomy}. As briefly mentioned in \S\ref{sec:methodology}, the multilevel configuration dependencies should be interpreted with the study methodology in mind, 
because they are derived from an incomplete set of configuration-related issues from two FS ecosystems. It is likely that there are more complex dependencies in  FS ecosystems, which deserves further investigation. 
 
\vspace{0.05in}
\noindent
{\bf Limitations of the \textsc{ConfD} framework}.
The current prototype 
requires a few user inputs (\S\ref{sec:annotation}) to guide the automated dependency analysis, which we hope to reduce  through more sophisticated state analysis.
The current taint analysis is dependent on configuration variables identified in the source code. We found that the configuration management in newer file and storage systems may not always maintain dedicated configuration variables for individual parameters, which may limit the scope of taint analysis.
Also,  \textsc{ConfD}  can only handle a subset of   LLVM  IR   for taint analysis and it only considers two parameters at a time for CPD and CCD, which may lead to incomplete dependency or false positives. 
We hope to improve these  through more advanced software engineering efforts in the future, which will likely improve the   effectiveness further. 

Similarly, there are limitations in  plugins. For example, 
\texttt{ConfD-handlingCk} only induces at most two violations for one configuration state for simplicity; there may be more issues if we consider more than two.
Also,
\texttt{ConfD-xfstests}   only transforms  a subset of the test suite due to the irregular configuration handling. 
{The failed test cases in \texttt{ConfD-xfstests} requires a lot of time and effort to further analyze the test case and get to the root cause of failure. This is unavoidable since this part requires deep file system internal knowledge to understand the issue. Although we have minimized the failure cases in our extension, it is still time consuming. In the future we would like to further minimize this by using AI to filter out some of the failed test cases to enhance efficiency.}

Despite the limitations, \prj has been effective in analyzing dependencies and exposing configuration-related issues in our experiments, so we believe that it  
will be valuable to the community.

\subsection{Future Work}
\noindent
{\bf Integration with other file systems and tools.}
As mentioned in Table~\ref{tab:manyFSandUtilities}, many file systems 
can be configured through different utilities, which could potentially benefit from the multilevel dependency analysis of \textsc{ConfD} after minor customization (e.g., providing  FS-specific inputs in JSON format~\ref{sec:annotation}).
Also, \textsc{ConfD} is complementary to  other modern tools besides the base tools used in   current plugins. 
For example, FB-HYDRA ~\cite{Yadan2019Hydra}    uses YAML files to store configurations which is compatible with the JSON files used by \textsc{ConfD}. Moreover, it supports a set of plugins called ``Sweepers'' to manipulate the selection of parameters. The dependency-based  state generation in \textsc{ConfD} could be implemented as one special ``Sweeper'' for FB-HYDRA~\cite{Yadan2019Hydra}. 
Similarly, the configurations generated by \textsc{ConfD} could potentially be integrated into CI/CD frameworks~\cite{Jenkins} to enable pipelined configuration-oriented testing and deployment. 

\vspace{0.05in}
\noindent
{\bf Support for other software}.
Configuration dependency is not limited to the  file systems (Ext4, XFS, ZFS) and  the storage engine (WiredTiger) studied in this paper.
For example, NDCTL~\cite{ndctl} is a utility to configure the \texttt{libnvdimm} subsystem in Linux.
We expect that adding NDCTL to the dependency analysis  will likely help address NVM-specific configuration issues more effectively.
Also, researchers   have observed functionality or correctness dependencies between local file systems and other software  
(e.g.,  databases~\cite{Zheng-OSDI14-DB}, distributed storage systems~\cite{Jinrui-ICS18,RedundancyDoesNotImplyFaultTolerance-FAST17,Runzhou-TOS22,Runzhou-PDSW20}), 
many of which are also related to  configurations. 
The  dependencies studied in this work   may serve as a foundation for investigating such configuration-related 
issues beyond file systems.
Also, since LLVM supports compiling a wide set of languages (e.g.,  C++, Rust, Swift) to IR through various frontends~\cite{LLVM},
the core analysis of \textsc{ConfD} is expected to be applicable to software written in other languages as well.  

\section{Conclusion}
\label{sec:conclusion}
 We have presented a study on 78 real-world configuration issues and built an extensible framework called \prj for addressing various configuration issues. Our experiments on Ext4, XFS, ZFS, and WiredTiger  demonstrate that \prj  can help address configuration issues effectively by leveraging configuration dependencies. In the future, we would like to improve \prj further and investigate other systems  as discussed in \S\ref{sec:discussion}.
 We hope that \prj can facilitate follow-up research on addressing the increasing  challenge of configurations in general.
 
\section*{Acknowledgments}
The authors would like to thank  the editor and
anonymous reviewers 
for their insightful feedback.  
We also thank Runzhou Han and Wei Xu for their help on validating and reproducing a few bug cases. This work was supported in part by National Science Foundation (NSF) under grants CNS-1855565, CCF-1853714, CCF-1910747  and CNS-1943204. Any opinions, findings, and conclusions expressed in this material are those of the authors and do not necessarily reflect the views of the sponsor. 

\printbibliography
\end{document}